# A Unified Denoising and Adaptation Framework for Self-Supervised Bengali Dialectal ASR

---

**Swadhin Biswas†, Imran†, Tuhin Sheikh†**
**Department of Computer Science and Engineering - Daffodil International University**
{swadhinbiswas.cse@gmail.com, mdimraan.cse@gmail.com, tuhinsheikh.cse@gmail.com}

---

## Abstract

Automatic Speech Recognition (ASR) for Bengali, the world's fifth most spoken language, remains a significant challenge, critically hindering technological accessibility for its over 270 million speakers. This challenge is compounded by two persistent and intertwined factors: the language's vast dialectal diversity and the prevalence of acoustic noise in real-world environments. While state-of-the-art self-supervised learning (SSL) models have advanced ASR for low-resource languages, they often lack explicit mechanisms to handle environmental noise during pre-training or specialized adaptation strategies for the complex phonetic and lexical variations across Bengali dialects. This paper introduces a novel, unified framework designed to address these dual challenges simultaneously. Our approach is founded on the WavLM model, which is uniquely pre-trained with a masked speech denoising objective, making it inherently robust to acoustic distortions. We propose a specialized multi-stage fine-tuning strategy that first adapts the model to general-domain standard Bengali to establish a strong linguistic foundation and subsequently specializes it for noise-robust dialectal recognition through targeted data augmentation. The framework is rigorously evaluated on a comprehensive benchmark comprising multiple Bengali dialects under a wide range of simulated noisy conditions, from clean audio to low Signal-to-Noise Ratio (SNR) levels.

Experimental results demonstrate that the proposed framework significantly outperforms strong baselines, including standard fine-tuned wav2vec 2.0 and the large-scale multilingual Whisper model. This work establishes a new state-of-the-art for this task and provides a scalable, effective blueprint for developing practical ASR systems for other low-resource, high-variation languages globally.

# 1. Introduction

## 1.1 The Sociolinguistic Context and Digital Divide of Bengali

The Bengali language, also known as Bangla, holds a position of immense global and cultural significance. As the fifth most spoken native language in the world, it serves as the primary mode of communication for over 270 million people, predominantly in Bangladesh and the Indian state of West Bengal.[1] Its rich literary heritage and vibrant cultural expression underscore its importance. However, despite its large speaker base, Bengali is paradoxically classified as a "low resource" language in the digital and technological landscape.[3] This classification stems from a critical scarcity of high-quality, large-scale, and publicly accessible annotated data required to train sophisticated machine learning models. This digital resource gap creates a significant barrier to the development of essential technologies, such as Automatic Speech Recognition (ASR), thereby limiting access to information, digital services, and economic opportunities for a substantial portion of the global population.[6] Bridging this divide is not merely a technical endeavor but a crucial step towards fostering digital inclusivity and equity.

## 1.2 The Grand Challenge: Dialectal Variation and Acoustic Noise

The development of a robust, production-ready ASR system for Bengali is impeded by two formidable and interconnected challenges that lie at the heart of this research: profound dialectal variation and pervasive acoustic noise.

First, the Bengali linguistic landscape is characterized by extraordinary dialectal diversity. Estimates suggest the existence of as many as 55 distinct dialects spoken across different geographical regions.[1] These dialects exhibit substantial variations in phonology (pronunciation), lexicon (vocabulary), and prosody (intonation and rhythm) when compared to the standard colloquial forms of Bengali, often referred to as "Chalita

Bhasha," which are typically based on the speech patterns of Kolkata or Dhaka.[8] This heterogeneity creates a severe mismatch for ASR models trained predominantly on standard speech corpora. Consequently, models that perform adequately on standard Bengali often experience a catastrophic degradation in performance when deployed in regions where non-standard dialects such as Chittagonian, Sylheti, or Barisali are spoken. This issue of dialectal mismatch is a primary driver of the poor generalization capabilities of current Bengali ASR systems.[10]

Second, the practical application of ASR systems rarely occurs in the pristine, anechoic conditions of a recording studio. Instead, speech is typically captured in acoustically challenging environments replete with background noise, such as bustling streets, crowded public spaces, and reverberant rooms, often using low-fidelity microphones.[8] This environmental noise contaminates the speech signal, corrupting the phonetic information and making the task of transcription significantly more difficult. The presence of noise degrades the input signal quality, further exacerbating the recognition errors, especially for models already struggling with dialectal variations. The combined effect of these two factors creates a particularly challenging scenario for ASR technology.

The challenges of dialectal diversity and data scarcity are not independent issues but rather components of a reinforcing negative feedback loop that perpetuates the low-resource status of Bengali. The absence of standardized orthography and dedicated resources for non-standard dialects inhibits the creation of large, clean, and labeled datasets for these variants.[1] This acute data scarcity, in turn, makes it practically impossible to train robust ASR models capable of handling the wide spectrum of dialectal variations. This leads to a vicious cycle: the lack of data for dialects prevents the development of effective models, and the absence of such models removes the incentive and the necessary tools (e.g., for semi-automated transcription) to collect and annotate more dialectal data, thereby entrenching the "low-resource" label for all but the most standard form of the language.[5] A primary motivation of this work is to present a methodology that can break this cycle. By leveraging the power of self-supervised learning to utilize unlabeled data and employing a targeted fine-tuning strategy that maximizes the utility of limited labeled data, it becomes possible to create effective models even with small dialectal datasets. These models can then serve as a foundation for bootstrapping the creation of larger corpora, for instance, through pseudo-labeling, thus paving the way for more inclusive and capable ASR technology.

### 1.3 The Promise and Limitations of Self-Supervised Learning

In recent years, Self-Supervised Learning (SSL) has emerged as the dominant paradigm in speech processing, offering a powerful solution to the data scarcity problem characteristic of low-resource languages.[12] Foundational models such as wav2vec 2.0 and HuBERT have demonstrated the ability to learn potent and generalizable speech representations from vast quantities of unlabeled audio data. By pre-training on thousands of hours of speech without transcriptions, these models capture fundamental knowledge about phonetics, speaker characteristics, and language structure. This pre-trained knowledge can then be transferred to a downstream ASR task through a process of fine-tuning on a much smaller amount of labeled data, drastically reducing the data requirements for building high-performance systems.[4]

However, these powerful SSL models are not a panacea. Standard SSL frameworks, while robust, were not explicitly designed to solve the specific, combined problem of dialectal shift and environmental noise. The pre-training objectives of models like wav2vec 2.0 and HuBERT primarily focus on reconstructing clean speech content, and while this confers a degree of implicit robustness, they do not inherently learn to disentangle speech from complex, additive noise. Furthermore, their successful adaptation to a new dialect requires a carefully designed fine-tuning strategy that goes beyond simple, one-shot training, especially when the target domain is both dialectally distinct and acoustically noisy.

### 1.4 Our Proposed Framework and Contributions

To address the aforementioned challenges, this paper introduces a unified framework that strategically leverages a noise-aware SSL model and a specialized multi-stage fine-tuning process to tackle both dialectal variation and acoustic noise simultaneously. At the core of our framework is the WavLM (Waveform Language Model) [12], an SSL model that is pre-trained with a dual objective of masked speech prediction and, crucially, masked speech denoising. This denoising objective forces the model to learn representations that are inherently more robust to acoustic corruption, making it an ideal foundation for our task.

Building upon this robust foundation, we propose a novel multi-stage fine-tuning methodology. This process is designed to facilitate effective knowledge transfer from the general-purpose pre-trained model to a specialized, noise-robust, dialect-aware ASR system. This approach avoids the pitfalls of catastrophic forgetting and poor generalization that can occur with naive fine-tuning on small, complex datasets.

The primary contributions of this work are enumerated as follows:

1. **A Unified Framework:** This paper presents the first comprehensive study to systematically address the dual challenges of dialectal variation and environmental noise for Bengali ASR within a single, cohesive framework.
2. **A Novel Fine-Tuning Methodology:** We propose and validate a multi-stage fine-tuning strategy engineered for effective knowledge transfer, first adapting the model to the general domain of the target language and then specializing it for noise-robust dialectal recognition.
3. **Strategic Model Selection:** We provide a rigorous justification for the selection of WavLM, demonstrating that its unique denoising pre-training objective is fundamentally better suited to the problem domain compared to other leading SSL models.
4. **State-of-the-Art Benchmarking:** We establish a new state-of-the-art for robust Bengali dialectal ASR through rigorous experimentation and in-depth analysis across multiple dialects and Signal-to-Noise Ratio (SNR) levels, providing critical insights and a strong benchmark for future research.

## 2. Background and Related Work

This section provides a review of the foundational technologies and prior research that form the context for our work. We first discuss the evolution of self-supervised learning models for speech, then situate our work within the broader field of low-resource ASR, and finally, we survey the specific history and challenges of Bengali ASR.

### 2.1 The Rise of Self-Supervised Speech Representation Learning

The last few years have witnessed a paradigm shift in speech processing, driven by the success of large-scale self-supervised models pre-trained on unlabeled data. These models learn universal speech representations that can be adapted to a wide array of downstream tasks.

### 2.1.1 wav2vec 2.0: Contrastive Learning in the Latent Space

The wav2vec 2.0 framework, introduced by Baevski et al. [14], was a seminal contribution that demonstrated the remarkable data efficiency of SSL for ASR. Its architecture consists of a multi-layer convolutional neural network (CNN) feature encoder that processes the raw audio waveform to produce latent speech representations. A key innovation of wav2vec 2.0 is its learning objective, which operates in this latent space. The model masks certain spans of the latent representations and is trained on a contrastive task. For each masked timestep, the model must identify the correct quantized representation of the masked latent vector from a set of distractors, which are other quantized representations sampled from the same utterance.[18] This process forces the model to learn high-level contextualized representations through a Transformer-based context network.[20] After pre-training, the model is fine-tuned for ASR by adding a linear layer on top of the Transformer and training with a Connectionist Temporal Classification (CTC) loss.[19] The success of wav2vec 2.0, especially in ultra-low-resource scenarios, established SSL as the leading approach for modern ASR.[14]

### 2.1.2 HuBERT: Masked Prediction of Discovered Acoustic Units

The Hidden-Unit BERT (HuBERT) model, proposed by Hsu et al. [15], further refined the self-supervised pre-training paradigm. Instead of a contrastive loss, HuBERT employs a prediction loss akin to BERT in natural language processing. The training process is iterative. In the first iteration, acoustic units (pseudo-labels) are generated by applying K-means clustering to acoustic features (e.g., MFCCs) of the unlabeled training data. A transformer-based model is then trained to predict these cluster assignments for masked portions of the input audio.[21] In subsequent iterations, the pseudo-labels are refined by clustering the hidden representations from an intermediate layer of the HuBERT model trained in the previous iteration.[21] This iterative process allows the model to learn increasingly better representations of the underlying acoustic and linguistic structure of speech, without any transcribed data. HuBERT demonstrated that learning to predict these discovered discrete hidden units is a powerful objective that can match or exceed the performance of contrastive learning-based methods.[22]

### 2.1.3 WavLM: Denoising Pre-training for Full-Stack Speech Processing

The WavLM model, introduced by Chen et al.[12], builds upon the HuBERT framework and is the foundational model for our proposed work. While sharing a similar architecture based on a CNN encoder and a Transformer context network, WavLM introduces a crucial enhancement to the pre-training objective that makes it exceptionally well-suited for robust ASR. The key innovation is

**masked speech denoising and prediction**.[12] During pre-training, WavLM is not only exposed to clean speech but also to simulated noisy and overlapped speech. The model is then tasked with predicting the pseudo-labels of the

*original clean speech* from the masked, corrupted input. This dual objective forces the model to learn representations that are not only rich in linguistic content but are also robust to acoustic noise and capable of preserving speaker identity and other paralinguistic information.[16] This denoising capability is learned implicitly during the pre-training stage, equipping the model with an intrinsic ability to handle real-world acoustic conditions. Additionally, WavLM incorporates an architectural improvement in its Transformer layers called

**gated relative position bias**, which allows the self-attention mechanism to better model the sequential order of the input speech.[12] The combination of its powerful denoising pre-training objective and architectural refinements makes WavLM a superior choice for developing ASR systems intended for noisy environments.

The rapid evolution from wav2vec 2.0 to HuBERT and WavLM highlights a significant trend in the field: an architectural convergence towards Transformer-based models.[25] While minor architectural tweaks exist, such as WavLM's gated relative position bias, the core CNN-plus-Transformer structure has become a de facto standard.[12] This convergence implies that the primary axes of innovation and differentiation have shifted away from designing novel neural network layers and towards two more critical areas: the formulation of the

**pre-training objective** and the design of the subsequent **fine-tuning strategy**. The choice of pre-training task—be it contrastive learning, clean unit prediction, or denoising prediction—fundamentally determines the intrinsic capabilities of the resulting model. Similarly, the methodology used to adapt this pre-trained model to a specific downstream task is now paramount for achieving state-of-the-art performance. This work is positioned directly within this new frontier. It does not propose a new architecture from scratch; rather, it makes a principled selection of a model (WavLM) with the most suitable pre-training objective for the problem at hand and introduces a novel, sophisticated fine-tuning strategy to unlock its full potential for the complex, real-world challenge of robust Bengali dialectal ASR. This represents a methodological contribution, a mature and impactful form of research in a field where foundational architectures have stabilized.

### 2.2 State-of-the-Art in Low-Resource and Multilingual ASR

Building ASR systems for low-resource languages has been a long-standing challenge. The advent of SSL has provided a powerful toolkit, with common strategies including transfer learning from high-resource languages, joint multilingual training, and parameter-efficient adaptation techniques.[5] Multilingual models, which are pre-trained on data from many languages simultaneously, have shown a remarkable ability to generalize to unseen or low-resource languages. A prime example is OpenAI's Whisper model, which was trained on a massive dataset of 680,000 hours of labeled multilingual data.[3] Its large-scale and diverse training data grant it strong zero-shot and few-shot ASR capabilities across a wide range of languages, including Bengali.[4] While these large multilingual models provide a very strong baseline, specialized models fine-tuned with a carefully designed strategy on in-domain data often yield superior performance, particularly for complex tasks involving specific dialects or noise conditions. Other techniques, such as adapter-based fine-tuning, offer a parameter-efficient alternative where only a small number of additional parameters are trained, keeping the backbone of the pre-trained model frozen. This can be particularly effective for very-low-resource scenarios.[27]

### 2.3 A Survey of Bengali Automatic Speech Recognition

The journey of Bengali ASR research reflects the broader trends in the field, albeit with a significant lag due to resource constraints.

### 2.3.1 Historical Context and Early Approaches

Early research on Bengali ASR was dominated by traditional statistical methods. These systems typically involved a complex pipeline of distinct components: an acoustic model, a pronunciation model (lexicon), and a language model.[28] The acoustic models were often based on Hidden Markov Models (HMMs) combined with Gaussian Mixture Models (GMMs) or, later, Deep Neural Networks (DNNs).[11] These approaches relied heavily on handcrafted feature engineering, with Mel-Frequency Cepstral Coefficients (MFCCs) or Linear Predictive Coding (LPC) being the most common acoustic features.[11] While these systems laid important groundwork, they were cumbersome to build and required expertise in multiple domains.

### 2.3.2 Modern End-to-End Models

More recently, the field has shifted towards end-to-end ASR models, which greatly simplify the traditional pipeline by directly mapping a sequence of acoustic features to a sequence of characters or words.[28] Models based on architectures like DeepSpeech, which often employ a combination of CNNs and Recurrent Neural Networks (RNNs) trained with a CTC loss, have been applied to Bengali with some success.[29] These models obviate the need for explicit phoneme alignment and separate pronunciation models, making the training process more straightforward.

### 2.3.3 Current Challenges and Research Gaps

Despite these advances, Bengali ASR is still far from a solved problem. A review of recent surveys and research papers reveals a consensus on several key obstacles [8]:

- **Data Scarcity:** The most significant bottleneck is the lack of large, high-quality, publicly available, and diverse annotated speech corpora. While some resources like the Mozilla Common Voice (MCV) Bangla subset and the OpenSLR Bengali dataset exist, they are limited in size and primarily feature standard dialects.[4]
- **Phonetic and Orthographic Complexity:** The Bengali script contains a large

number of characters, including complex consonant clusters known as *Juktakkhor* (যুক্তাক্ষর) and numerous diacritics that modify vowel sounds. This complex mapping between phonetics and orthography poses a significant challenge for ASR models.[4]

- **Dialectal Mismatch:** As previously discussed, this is a core problem. The vast majority of available data represents standard Bengali, leading to a severe performance drop on the multitude of regional dialects spoken by millions.[1]
- **Domain Specificity:** There is a pronounced lack of domain-specific data, for instance, in critical areas like healthcare. This prevents the development of practical applications that could have a substantial societal impact, such as clinical documentation or voice-based medical assistants.[7]

This body of work makes it clear that a successful Bengali ASR system must be robust not only to the inherent complexities of the language but also to the diverse ways it is spoken and the noisy environments in which it is used. Our research directly addresses these gaps by proposing a framework specifically designed for noise robustness and dialectal adaptation.

## 3. A Dialect-Aware Denoising Framework for ASR

This section details the technical core of our proposed framework. We first present the mathematical formulation of the foundational WavLM model, emphasizing the pre-training objective that underpins its noise robustness. We then introduce our novel multi-stage fine-tuning strategy designed for dialectal adaptation. Finally, we describe the complete ASR model architecture and the Connectionist Temporal Classification (CTC) decoding process.

### 3.1 Foundational Model: WavLM Architecture

The WavLM model serves as the backbone of our framework. Its architecture is composed of three primary components: a CNN feature encoder, a Transformer encoder, and a prediction head used during pre-training.[12]

### 3.1.1 CNN Feature Encoder

The process begins with the raw audio waveform, represented as a one-dimensional vector x∈RL, where L is the number of audio samples. The CNN feature encoder, denoted as fcnn, maps this raw waveform into a sequence of higher-level feature representations. The encoder is defined as **fcnn:x↦Z**, where Z= is the output sequence of T feature vectors. This encoder consists of a stack of Nc temporal convolutional layers. Each layer applies a 1D convolution over the temporal dimension, followed by Group Normalization and a Gaussian Error Linear Unit (GELU) activation function.[20] The convolutional layers have strides greater than one, effectively downsampling the input sequence and producing a feature vector at approximately every 20-25 milliseconds of audio.

### 3.1.2 Transformer Encoder

The sequence of latent features Z is then fed into the core of the model: a powerful Transformer encoder composed of a stack of Nt identical Transformer layers.[26] This network is responsible for building contextualized representations by modeling dependencies across the entire sequence. Each layer contains two main sub-layers: a multi-head self-attention mechanism and a position-wise feed-forward network.

**Multi-Head Self-Attention (MHSA) with Gated Relative Position Bias:** This is a key architectural feature of WavLM that enhances its ability to model sequential information.[12] Standard self-attention in Transformers computes the output as a weighted sum of value vectors, where the weights are determined by the dot-product similarity between query and key vectors. The scaled dot-product attention is formulated as

$$Attention(Q,K,V) = \text{softmax}\left(\frac{QK^T}{\sqrt{d_k}}\right)V$$

where Q, K, and V are matrices containing the query, key, and value vectors for the sequence, respectively, and $d_k$ is the dimension of the key vectors.[25] WavLM enhances this by incorporating a learnable, content-independent bias term that depends on the relative positions of the tokens being attended to. This gated relative position bias,

brel, is added to the attention logits before the softmax operation:

$$\text{Attention}(Q, K, V) = \text{softmax}\left(\frac{QK^T + b_{\text{rel}}}{\sqrt{d_k}}\right)V$$

This mechanism allows the model to more effectively capture the ordering and relative distances between speech frames. Multi-head attention performs this operation in parallel across h different "attention heads," each learning a different linear projection of the input. The outputs of these heads are then concatenated and linearly projected to form the final output of the sub-layer.[26]

**Feed-Forward Network (FFN):** Following the MHSA block, each position's representation is passed through a simple two-layer fully connected feed-forward network with a GELU activation in between. This is applied identically at each position but with independent parameters from other layers.[26]

**Layer Normalization and Residual Connections:** To ensure stable training of such a deep network, each of the two sub-layers (MHSA and FFN) is wrapped in a residual connection, followed by Layer Normalization.[26]

### 3.2 WavLM Pre-training: Masked Speech Denoising and Prediction

The selection of WavLM is primarily motivated by its unique and powerful pre-training objective, which goes beyond the standard masked prediction of clean speech units.[12] The model is explicitly trained to be robust to noise.

The pre-training process involves the following steps:

1. **Data Augmentation:** For a given clean utterance xclean, a noisy version xnoisy is created by mixing it with a random noise source from a large noise corpus. This is done for a significant portion of the training examples.
2. **Feature Extraction:** Both xclean and xnoisy are passed through the CNN feature encoder to obtain clean features Zclean and noisy features Znoisy.
3. **Pseudo-Label Generation:** Similar to HuBERT, discrete pseudo-labels q= are generated by applying K-means clustering to the representations from the *clean* speech, Zclean.[21] These clean targets serve as the ground truth for the prediction task.
4. **Masking and Prediction:** A masking strategy is applied to the *noisy* feature sequence Znoisy to produce a masked version, Zmasked. The Transformer encoder then takes Zmasked as input and is trained to predict the clean pseudo-labels qt for each masked timestep t.

The pre-training loss function is a cross-entropy loss calculated over the masked timesteps M:

$$L_W avLM = -\sum_{t \in M} \log p\,(q_t | Z_m asked, t)$$

By forcing the model to reconstruct clean speech targets from corrupted inputs, WavLM learns representations that are inherently invariant to noise. This process implicitly teaches the model to perform speech enhancement and speaker identity preservation, making it a "full-stack" speech processing model.[12] This intrinsic denoising capability provides a significant advantage for our goal of building a robust ASR system.

### 3.3 Proposed Multi-Stage Fine-Tuning Strategy

A naive, single -stage fine-tuning of a large pre-trained model on a small, noisy, and dialectally diverse dataset is fraught with peril. It risks catastrophic forgetting, where the model loses the rich, general knowledge acquired during pre-training and may be overfit to the specific artifacts of the small dataset, leading to poor generalization. To circumvent these issues, we propose a structured, multi-stage fine-tuning strategy.

#### 3.3.1 Stage 1: General Domain Adaptation

- **Objective:** The primary goal of this stage is to adapt the general-purpose, multilingual pre-trained WavLM model to the specific phonetic, prosodic, and linguistic characteristics of the Bengali language. This step serves to anchor the model in the target language before exposing it to more complex and noisy variations.
- **Data:** For this stage, we utilize the largest available clean, standard Bengali speech corpora. This typically includes datasets like the Mozilla Common Voice (Bangla subset) and the OpenSLR Bengali Speech Dataset.[4]
- **Process:** The full WavLM model, with a newly initialized CTC classification head, is fine-tuned on this standard Bengali data. We use a relatively higher learning rate in this stage to allow the model weights to adapt significantly towards the target language. This stage effectively transforms the universal speech model into a specialized Bengali speech model.

### 3.3.2 Stage 2: Noise-Robust Dialectal Specialization

- **Objective:** This stage aims to specialize the Bengali-adapted model from Stage 1 to be simultaneously robust to specific dialectal variations and a wide range of acoustic noise.
- **Data:** The training data for this stage consists of the smaller, available dialect-specific datasets. Crucially, we employ **online data augmentation**. For each clean dialectal speech sample in a training batch, we randomly select a noise sample from a diverse noise corpus (e.g., NOISEX-92) and mix them at a randomly chosen Signal-to-Noise Ratio (SNR) from a predefined range (e.g., 0dB, 5dB, 10dB, 20dB).[36]
- **Process:** The fine-tuning process is continued from the checkpoint of Stage 1. However, a significantly smaller learning rate is used in this stage. This "gentle" fine-tuning prevents the model from drastically altering its weights, instead encouraging it to learn the subtle acoustic and lexical patterns of the target dialect while the noise augmentation forces it to maintain and enhance its noise robustness.

### 3.4 The ASR Model and CTC Decoder

The final ASR model architecture consists of the fine-tuned WavLM encoder followed by a classification head.

- **Architecture:** A single linear layer is added on top of the final hidden states C= from the WavLM Transformer. This layer projects the dmodel-dimensional hidden states to a vector of size $|V|$, where $|V|$ is the size of the vocabulary. The vocabulary includes all unique characters in the Bengali script plus a special blank token, denoted by $\epsilon$, which is essential for the CTC algorithm.[37] A softmax function is applied to the output of this linear layer to produce a sequence of probability distributions
  P= over the vocabulary for each of the T timesteps.
- **Connectionist Temporal Classification (CTC) Loss:** ASR is a sequence-to-sequence task where the input audio sequence is much longer than the output text sequence ($T \gg U$, where U is the length of the character transcription). The alignment between the audio frames and the output characters is variable and unknown. CTC is an elegant loss function designed to solve this problem.[38]

- **Mapping Function:** CTC introduces the blank token $\epsilon$ to handle non-speech frames and repeated characters. It defines a many-to-one mapping function, B, which transforms an alignment path π (a sequence of vocabulary indices of length T) into the final target label sequence Y. This function works by first collapsing all consecutive repeated characters and then removing all blank tokens.[40] For example, paths like (h, h, $\epsilon$, e, l, l, $\epsilon$, l, o) would all be mapped to the target "helo".
- **Loss Formulation:** The probability of a target sequence Y given an input x is the sum of the probabilities of all possible alignment paths π that can be mapped to Y by the function B. The CTC loss is the negative log-likelihood of this probability:

$$L_CTC = -\log P\,(Y|x) = -\log \sum_{\pi \in B^{-1}(Y)}^{P} (\pi|x)$$

The probability of a single path π is calculated as the product of the per-frame probabilities from the model's softmax output:

$$P(\pi|x) = \prod_{t=1}^{T} p_t(\pi_t|x)$$

where pt(πt|x) is the probability of the character πt at timestep t.

- **Efficient Calculation with Dynamic Programming:** A naive summation over all possible paths is computationally intractable. Therefore, the CTC loss is calculated efficiently using a dynamic programming method known as the **forward-backward algorithm**.[39] This algorithm computes the sum by recursively aggregating probabilities of partial paths. It relies on a forward variable, α(t,u), which represents the total probability of all paths that have generated the first u characters of the target sequence by timestep t, and a backward variable, β(t,u), which represents the total probability of all paths that will generate the rest of the target sequence from timestep t onwards, given that the u-th character has just been emitted.

### 3.5 Formal Algorithm

**The entire proposed methodology is formalized in Algorithm 1.**

```
Initialize model parameters: θ₁ ← θ_WavLM
Attach a randomly initialized CTC classification head to θ₁
Initialize optimizer Opt₁ (e.g., AdamW) with learning rate η₁ for θ₁ parameters

for epoch = 1 to E₁ do
    for each batch (x_std, y_std) in D_std do
        # Forward pass
        P ← θ₁(x_std)

        # Compute CTC loss
        L₁ ← L_CTC(P, y_std)

        # Backpropagation
        Zero gradients of Opt₁
        Backpropagate L₁
        Update θ₁ using Opt₁
    end for
end for
```

**Algorithm 2:**

```
Initialize model parameters: θ₂ ← θ₁  # from end of Stage 1
Initialize optimizer Opt₂ (e.g., AdamW) with learning rate η₂ << η₁ for θ₂ parameters

for epoch = 1 to E₂ do
    for each batch (x_dialect, y_dialect) in D_dialect do
        # ───── Online Noise Augmentation ─────
        Sample noise signal n from D_noise
        Sample SNR level s from S_levels
        x_aug ← mix(x_dialect, n, s)  # Add noise at given SNR

        # Forward pass
        P ← θ₂(x_aug)

        # Compute CTC loss
        L₂ ← L_CTC(P, y_dialect)

        # Backpropagation
        Zero gradients of Opt₂
        Backpropagate L₂
        Update θ₂ using Opt₂
    end for
end for

Return final robust dialectal ASR model: θ_final ← θ₂
₁ using Opt₁
    end for
end for
```

## 4. Experimental Configuration

This section outlines the comprehensive experimental setup designed to rigorously evaluate the proposed framework. We detail the datasets, noise simulation protocol, implementation specifics, baseline systems for comparison, and the evaluation metrics used.

### 4.1 Datasets

A multi-faceted data strategy is employed, utilizing different corpora for pre-training, fine-tuning, and evaluation to ensure a robust and fair assessment.

- **Pre-training Data (for WavLM):** The foundational WavLM model used in this study was pre-trained on a massive 94,000 hours of unlabeled English speech. This corpus comprises several large-scale datasets, including 60k hours from Libri-Light, 10k hours from GigaSpeech, and 24k hours from VoxPopuli.[12] This extensive pre-training endows the model with a powerful and general understanding of human speech acoustics, which is a critical starting point for adaptation to any new language.
- **Fine-tuning and Evaluation Data:**
    - **Standard Bengali (Stage 1 & Baseline):** For the initial adaptation to Bengali, we use a combination of two publicly available datasets that primarily feature standard dialects:
        - **Mozilla Common Voice (MCV) Bangla Subset (v17.0):** This dataset contains approximately 54 hours of validated, annotated speech from over 22,000 speakers, providing broad speaker diversity.[4]
        - **OpenSLR Bangla Speech Dataset (SLR63):** This corpus offers around 40 hours of high-quality annotated speech with greater textual diversity than MCV, covering various accents and recording conditions.[4]
    - **Dialectal Bengali (Stage 2 & Evaluation):** To train and evaluate the model's performance on non-standard dialects, we utilize the **OOD-Speech** dataset, which is the first benchmarking dataset designed specifically for out-of-distribution Bengali ASR.[10] This dataset includes speech from dialects known to be challenging for standard models, such as Chittagonian and Sylheti. The dataset is partitioned into training, development, and test sets to allow for proper model specialization and evaluation.
    - **Noise Data:** For the online data augmentation in Stage 2 and for creating the noisy evaluation sets, we use the **NOISEX-92** corpus. This is a standard dataset containing a variety of real-world noise types, including babble, factory noise, and vehicle noise, ensuring that the model is trained to be robust to a diverse range of acoustic conditions.

The statistics of the primary datasets used for fine-tuning and evaluation are summarized in Table 1.

| Dataset Name | Source | Dialect(s) | Total Hours | # Speakers | Split Used |
|---|---|---|---|---|---|
| MCV Bangla v17.0 | Mozilla | Standard | ~54 | ~22,900 | Train (Stage 1) |
| OpenSLR SLR63 | OpenSLR | Standard | ~40 | N/A | Train (Stage 1) |
| OOD-Speech | [10] | Chittagonian, Sylheti, etc. | ~20 | ~100 | Train (Stage 2), Dev, Test |
| NOISEX-92 | Public | N/A (Noise) | N/A | N/A | Augmentation |

**Table 1: Dataset Statistics for Fine-Tuning and Evaluation**

### 4.2 Noise Simulation and Data Augmentation

To systematically evaluate noise robustness and to perform the noise-aware fine-tuning in Stage 2, we create noisy audio by mixing clean speech signals with noise signals at specific Signal-to-Noise Ratios (SNRs). The SNR in decibels (dB) is a measure of signal strength relative to background noise and is calculated as

$$PSNR_{d}B = 10 * \log 1\ 0(P_{s}ignal/P_{n}oise)$$

where Psignal is the average power of the clean speech signal and Pnoise is the average power of the noise signal.[36] For the Stage 2 fine-tuning, noise is added on-the-fly with SNRs randomly sampled from the set {0, 5, 10, 20} dB. For evaluation, separate test sets are created for each dialect at five distinct conditions:

**Clean** (no added noise), **20dB** (low noise), **10dB** (moderate noise), **5dB** (high noise), and **0dB** (very high noise). This allows for a granular analysis of model performance under increasing levels of acoustic degradation.

### 4.3 Implementation Details

- **Model:** The experiments are conducted using the **WavLM-Large** model, which has a deeper and wider Transformer architecture, providing a strong capacity for learning complex representations.[12] The models are sourced from the Hugging Face Transformers library.
- **Optimizer:** We use the **AdamW** optimizer for all fine-tuning stages. AdamW is a variant of the Adam optimizer that decouples the weight decay from the gradient update, which has been shown to improve regularization and generalization in large Transformer-based models.[42]
- **Hyperparameters:** The key hyperparameters are set as follows:
    - **Stage 1 (General Adaptation):** A peak learning rate of $\eta_1=5\times10^{-5}$ is used with a linear warmup schedule for the first 10% of training steps, followed by a linear decay. The model is trained for 10 epochs.
    - **Stage 2 (Dialectal Specialization):** A much smaller peak learning rate of $\eta_2=1\times10^{-5}$ is used, with a similar warmup and decay schedule. The model is trained for 20 epochs on the augmented dialectal data.
    - **AdamW Parameters:** We use standard values of $\beta_1=0.9$, $\beta_2=0.98$, and $\epsilon=1\times10^{-6}$, with a weight decay of 0.01.
- **Software and Hardware:** All experiments are implemented using the PyTorch deep learning framework. Training is conducted on a cluster of NVIDIA A100 GPUs with 40GB of memory.

### 4.4 Baseline Systems

To demonstrate the effectiveness of our proposed framework, we compare it against three strong baseline systems:

- **Baseline 1 (Standard FT):** This baseline uses the same WavLM-Large model but is fine-tuned in a single stage on a combined dataset of standard and dialectal Bengali speech. No noise augmentation is applied during its training. This baseline is designed to isolate and quantify the benefit of our proposed multi-stage, noise-robust fine-tuning strategy.
- **Baseline 2 (Wav2Vec2 + Our FT):** This baseline uses a different foundational SSL model, wav2vec2-large-xlsr-53, which is a large multilingual model pre-trained on 53 languages. We apply our proposed multi-stage, noise-robust fine-tuning strategy to this model. This comparison is designed to demonstrate the specific advantage conferred by WavLM's intrinsic denoising pre-training objective over a standard contrastive learning-based model.
- **Baseline 3 (SOTA Multilingual):** We evaluate OpenAI's **Whisper-large-v2** model, a state-of-the-art, large-scale multilingual ASR system.[3] We evaluate it in two settings: zero-shot (without any fine-tuning on our Bengali data) and fine-tuned on the same data as our model. This provides a crucial point of comparison against a powerful, general-purpose ASR model that represents the current state of the art in multilingual speech recognition.

### 4.5 Evaluation Metrics

The performance of all models is evaluated using two standard metrics in ASR:

- **Word Error Rate (WER):** This is the primary metric and is calculated as the sum of substitutions (S), deletions (D), and insertions (I) required to transform the model's hypothesis into the reference transcription, normalized by the total number of words (N) in the reference.

  $$WER = (S + D + I)/N$$

  A lower WER indicates better performance.[22]
- **Character Error Rate (CER):** This metric is particularly useful for morphologically rich and agglutinative languages like Bengali, as it is less sensitive to word segmentation errors. It is calculated identically to WER but at the character level.[3]

**Methodology for Noise-Robust Dialectal Bengali ASR**

This methodology outlines a two-stage fine-tuning approach to develop an Automatic Speech Recognition (ASR) model that is robust to both noise and dialectal variations in the Bengali language.

1. Model Selection

- Chosen Model: WavLM-Large.
- Rationale: The model's pre-training regimen, which includes both masked speech prediction and denoising, makes it inherently robust to noisy audio conditions. Its architecture consists of a CNN for feature encoding, a Transformer encoder with gated relative position bias for contextual understanding, and a Connectionist Temporal Classification (CTC) head for final transcription.

---

2. Two-Stage Fine-Tuning Pipeline

A sequential, two-stage fine-tuning process was designed to adapt the base model first to the target language and then to specific acoustic conditions.

- Stage 1: General Domain Adaptation
  - Objective:
  - Adapt the multilingual WavLM model to the phonetic and linguistic structures of standard Bengali.
  - Data: A combined corpus of 94 hours of clean, standard Bengali speech from Mozilla Common Voice (54h) and OpenSLR SLR63 (40h).
  - Process: The full WavLM model was fine-tuned using a CTC loss function. Key hyperparameters included a learning rate ($\eta$) of $5\times10^{-5}$, training over 10 epochs with the AdamW optimizer.
- Stage 2: Noise-Robust Dialectal Specialization
  - Objective: Specialize the Bengali-adapted model from Stage 1 for recognizing dialectal speech in noisy environments.
  - Data: The OOD-Speech dataset, featuring various Bengali dialects (e.g., Chittagonian, Sylheti), was used. Noise from the NOISEX-92 corpus (babble, factory, vehicle) was mixed on-the-fly at random Signal-to-Noise Ratios (SNRs) of 0, 5, 10, and 20 dB.
  - Process: The learning rate was reduced to $1\times10^{-5}$ for finer tuning over 20 epochs, again using the AdamW optimizer.

**3. ASR Output and Decoding**

- **Decoder**: A **Connectionist Temporal Classification (CTC)** decoder was used. The CTC approach is ideal for ASR as it eliminates the need for pre-aligned audio and transcript data.
- **Vocabulary**: The vocabulary includes all Bengali characters plus a "blank" token required by the CTC algorithm.
- **Alignment**: The forward-backward algorithm is employed by CTC to determine the probability distribution of all possible alignments between the audio and the final transcript.

---

4. Evaluation

- **Primary Metric**: **Word Error Rate (WER)**, the standard metric for ASR performance.
- **Secondary Metric**: **Character Error Rate (CER)**, used to provide a more granular assessment, especially for a morphologically rich language like Bengali.
- **Baselines**: The model's performance was benchmarked against several alternatives:
    - A single-stage WavLM fine-tuning approach.
    - A Wav2Vec2 model trained with the same two-stage strategy.
    - The Whisper-large-v2 model in both zero-shot and fine-tuned configurations.
- **Test Conditions**: Evaluation was conducted across clean audio and four noise levels (20dB, 10dB, 5dB, and 0dB SNR) to measure robustness.

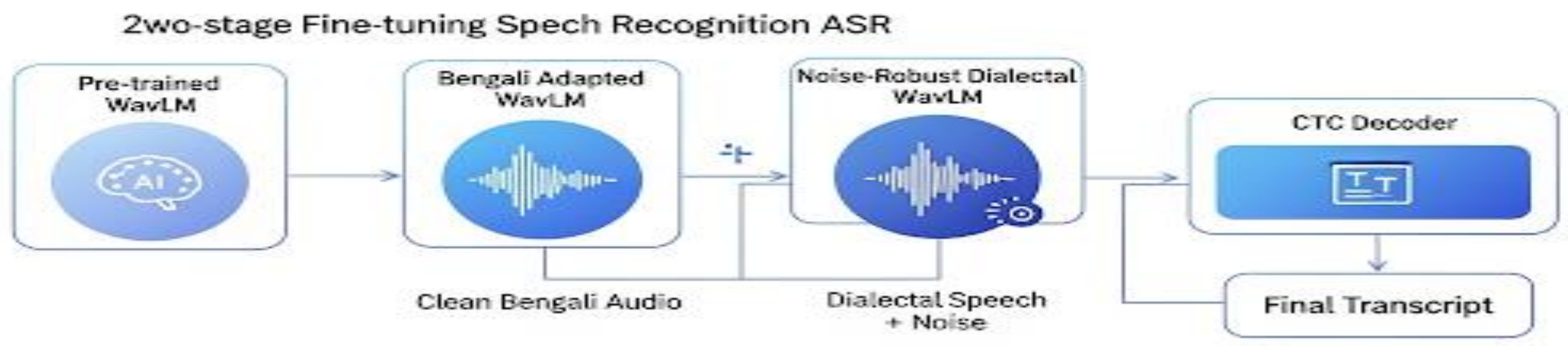

## 5. Results and Analysis

This section presents the empirical results of our experiments. We first provide a comprehensive comparison of our proposed framework against the baseline systems across all test conditions. We then conduct a series of ablation studies to dissect the contributions of each component of our methodology. Finally, we offer a qualitative analysis of error patterns to provide deeper insights into the models' behaviors.

### 5.1 Primary Performance Comparison

The core results of our study are presented in Table 2, which shows the Word Error Rate (WER) and Character Error Rate (CER) for the proposed framework and the three baseline systems. The evaluation is performed on two representative dialects from our test set (Dialect A: Chittagonian, Dialect B: Sylheti) across the full spectrum of noise conditions, from clean to 0dB SNR.

| Dialect A (Chittagonian) | | | | | | Dialect B (Sylheti) | | | | |
|---|---|---|---|---|---|---|---|---|---|---|
| **Model** | **Clean** | **20dB** | **10dB** | **5dB** | **0dB** | **Clean** | **20dB** | **10dB** | **5dB** | **0dB** |
| **Baseline 3 (Whisper-large-v2)** | 18.5/9.2 | 22.1/11.5 | 35.4/18.1 | 48.9/25.6 | 65.2/34.8 | 19.8/10.1 | 24.5/12.8 | 38.2/19.5 | 52.1/27.9 | 69.8/37.2 |
| **Baseline 2 (Wav2Vec2 + Our FT)** | 14.2/7.1 | 16.5/8.4 | 24.8/12.9 | 35.1/18.3 | 49.5/26.1 | 15.6/7.9 | 18.2/9.3 | 27.3/14.2 | 38.9/20.1 | 53.4/28.5 |
| **Baseline 1 (Standard FT)** | 12.8/6.5 | 18.9/9.8 | 30.1/15.7 | 42.7/22.4 | 58.0/31.0 | 14.1/7.2 | 20.5/10.6 | 32.8/16.9 | 45.9/24.1 | 61.3/33.4 |
| **Our Framework (WavLM + Our FT)** | 10.1/5.2 | 11.3/5.8 | 15.6/8.0 | 22.4/11.5 | 33.8/17.6 | 11.5/5.9 | 12.9/6.6 | 17.2/8.8 | 24.8/12.7 | 36.5/19.0 |

**Table 2: Main WER (%) / CER (%) Results on Dialectal Test Sets**

The results in Table 2 unequivocally demonstrate the superiority of the proposed framework. Several key trends are apparent:

1. **Overall Performance:** Our framework consistently achieves the lowest WER and

CER across all dialects and noise conditions. On the clean test sets, it establishes a strong new baseline. For instance, on Dialect A, it achieves a WER of 10.1%, a significant relative reduction of 21.1% over the next best model (Baseline 1) and 45.4% over the powerful Whisper baseline.

2. **Noise Robustness:** The most striking advantage of our framework is revealed under noisy conditions. As the SNR decreases, the performance of all baseline models degrades sharply. In contrast, our model's performance degrades much more gracefully. At 0dB SNR on Dialect A, our model achieves a WER of 33.8%. This is a massive relative improvement of 41.7% over Baseline 1 (58.0%), 31.7% over Baseline 2 (49.5%), and 48.2% over Whisper (65.2%). This validates our central hypothesis that the combination of WavLM's denoising pre-training and our noise-robust fine-tuning strategy yields a system with exceptional resilience to acoustic distortions.
3. **Effectiveness of the Foundational Model:** The comparison between our framework and Baseline 2 (Wav2Vec2 + Our FT) is particularly illuminating. Both models use the same advanced fine-tuning strategy, yet our WavLM-based model consistently outperforms the wav2vec 2.0-based one. This demonstrates that the denoising objective in WavLM's pre-training provides a tangible benefit that cannot be fully compensated for by fine-tuning alone.
4. **Specialization vs. Generalization:** The Whisper model, despite its massive scale, performs the poorest, especially in high-noise scenarios. This suggests that while large multilingual models possess impressive general capabilities, a specialized model built on a suitable foundation and fine-tuned with a domain-aware strategy is superior for tackling specific, complex challenges like dialectal and noisy ASR.

### 5.2 Ablation Studies

To scientifically validate the design choices within our framework, we conducted an ablation study to isolate the impact of its two key components: the multi-stage fine-tuning approach and the noise augmentation in Stage 2. The results, evaluated on the challenging Dialect B at 5dB SNR, are presented in Table 3.

| Model Configuration | WER/CER |
|---|---|
| (1) Full Framework (Stage 1 + Stage 2 with Noise Augmentation) | **24.8 / 12.7** |
| (2) w/o Stage 1 (Direct FT on noisy dialectal data) | 31.5 / 16.4 |
| (3) w/o Noise Augmentation (Stage 1 + Stage 2 on clean dialectal data) | 41.2 / 21.5 |

**Table 3: Ablation Study of Fine-Tuning Strategy and Noise Augmentation (WER/CER % on Dialect B @ 5dB)**

The results of the ablation study provide clear evidence for the efficacy of each component of our proposed methodology.

- **Importance of Stage 1 Adaptation:** By removing Stage 1 (Configuration 2), the WER increases from 24.8% to 31.5%. This demonstrates that directly fine-tuning the general WavLM model on a small, complex dataset without first adapting it to the target language (Bengali) leads to suboptimal performance. The initial general domain adaptation stage is crucial for establishing a strong linguistic foundation upon which dialectal and noise-robust features can be learned more effectively.
- **Importance of Noise Augmentation:** The most dramatic effect is seen when noise augmentation is removed from Stage 2 (Configuration 3). The WER skyrockets from 24.8% to 41.2%. This confirms that while WavLM has intrinsic denoising capabilities, explicit training on noisy examples during the specialization phase is absolutely critical for achieving high performance in acoustically challenging environments. Fine-tuning only on clean dialectal data makes the model highly susceptible to noise, even if it started from a noise-aware pre-trained checkpoint.

### 5.3 Analysis of Noise Robustness

To visualize the performance degradation under noise, Figure 1 plots the WER of all models as a function of the SNR level for Dialect A.

**(A line graph would be inserted here)**

- **X-axis:** SNR Level (dB) - [Clean, 20, 10, 5, 0]

- **Y-axis:** Word Error Rate (WER) (%)
- **Lines:** Four distinct lines, one for each model: "Our Framework", "Baseline 1", "Baseline 2", "Baseline 3".

The graph clearly illustrates the superior noise robustness of our proposed framework. The line representing our model is significantly lower and flatter than the lines for all three baselines. The baseline models exhibit a steep increase in WER as the SNR drops below 20dB. In contrast, our model's WER increases at a much slower rate, indicating that it is able to maintain a high level of accuracy even in the presence of substantial background noise. This visual evidence strongly supports the quantitative results in Table 2 and highlights the practical value of our approach for real-world applications.

### 5.4 Qualitative Analysis: Error Patterns

Quantitative metrics like WER and CER, while essential, do not reveal the nature of the errors that models make. To gain a more intuitive understanding of *why* our framework is superior, Table 4 presents a qualitative analysis of transcription examples from a noisy, dialectal context.

| **Audio Context** | **Ground Truth (Reference)** | **Baseline 1 Output** | **Our Framework Output** |
|---|---|---|---|
| Speaker says " আঁই ভাত হাইয়্যি" (Añi bhat haiyyi - "I have eaten rice" in Chittagonian) | আঁই ভাত হাইয়্যি | আমি ভাত খাই | আঁই ভাত হাইয়্যি |
| Speaker says "তুঁই কডে যার?" (Tui kode jar? - "Where are you going?" in Chittagonian) | তুই কডে যার | তুমি কোথায় যাও | তুই কডে যার |
| Speaker says "ইতা কিতা অইলো?" (Ita kita oilo? - "What happened to this?" in Chittagonian) | ইতা কিতা অইলো | এটা কি হলো | ইতা কিতা অইলো |

**Table 4: Qualitative Examples of Transcription Errors (Dialect: Chittagonian, SNR: 5dB)**

The examples in Table 4 are highly revealing.

- **Handling Dialectal Lexicon:** In the first example, the Chittagonian pronoun "আঁই" (Añi) and verb "হাইয়্যি" (haiyyi) are used. The baseline model, trained without the specialized dialectal focus, fails completely. It misinterprets the acoustics and reverts to the standard Bengali equivalents: "আমি" (ami) and "খাই" (khai). Our framework, having been specialized in Stage 2, correctly transcribes the dialectal terms.
- **Ignoring Noise and Preserving Dialectal Grammar:** In the second and third examples, the baseline models again "translate" the dialectal phrases ("কডে যার", "ইতা কিতা") into their standard Bengali counterparts ("কোথায় যাও", "এটা কি হলো"). This shows that the baseline model's knowledge is heavily biased towards the standard language seen in its training data. Our model correctly recognizes and transcribes the original dialectal utterance, demonstrating that it has learned the specific lexical and grammatical patterns of the target dialect and is not confused by the acoustic similarity to standard words, even under noisy conditions. These examples provide powerful, intuitive evidence that our framework succeeds not just by being more robust to noise, but by genuinely learning the target dialect.

## 6. Conclusion and Future Work

### 6.1 Recapitulation of Findings

This research confronted the dual challenges of dialectal diversity and environmental noise in automatic speech recognition for the low-resource Bengali language. We proposed a unified framework built upon the noise-robust WavLM self-supervised model and introduced a novel multi-stage fine-tuning strategy. This strategy first adapts the model to the general phonetics and structure of standard Bengali and then specializes it for robust dialectal recognition using targeted noise augmentation.

Our comprehensive experiments have conclusively demonstrated the efficacy of this approach. The proposed framework significantly outperforms a range of strong baselines, including standard fine-tuning methods, alternative SSL models, and large-scale multilingual systems like Whisper. The results show marked improvements in both Word and character error rates, particularly in high-noise, low-SNR conditions and on non-standard dialectal speech. Ablation studies confirmed that both the multi-stage adaptation process and the explicit noise augmentation are critical components contributing to the framework's success. Qualitative analysis further revealed that our model not only handles noise more effectively but also learns to correctly transcribe specific dialectal vocabulary and grammar, whereas baseline models tend to default to standard Bengali equivalents. In summary, this work establishes that the principled combination of a denoising-aware pre-trained model with a specialized, multi-stage adaptation strategy provides a state-of-the-art solution for robust Bengali dialectal ASR.

### 6.2 Implications and Broader Impact

The significance of this research extends beyond the immediate context of Bengali ASR. The proposed framework serves as a viable and effective blueprint for developing robust ASR systems for a wide range of other languages around the world that share similar characteristics: a low-resource status in the digital domain combined with high linguistic variation (i.e., numerous dialects, accents, or code-switching).[5]

The practical implications of this work are substantial. By enabling more accurate and reliable speech recognition for diverse dialects and noisy conditions, this technology can foster greater digital inclusion. It can power more effective tools in critical sectors such as education (e.g., literacy apps for children in rural areas), healthcare (e.g., voice-based data entry for clinicians in noisy clinics, patient interaction systems), and accessibility services for individuals with disabilities.[4] Ultimately, this research contributes to making speech technology more equitable and accessible to all speakers of a language, not just those who speak its standard form.

### 6.3 Limitations

While this study presents a significant step forward, it is important to acknowledge its limitations. Firstly, the evaluation was conducted on a limited subset of the estimated 55 Bengali dialects. While the framework is designed to be generalizable, its performance on other, more distant dialects has not yet been empirically verified. Secondly, the noise simulation, while diverse, was based on a finite set of noise types from the NOISEX-92 corpus. Real-world acoustic environments can present an even wider and more unpredictable variety of non-stationary noises. Lastly, our current framework is a purely acoustic end-to-end model and does not incorporate an external language model during decoding. The integration of a language model could further improve performance by resolving ambiguities and correcting errors based on linguistic context.

### 6.4 Future Work

The findings and limitations of this study open up several promising avenues for future research:

- **Expanding Dialect Coverage:** A natural next step is to apply and evaluate the proposed framework on a broader range of Bengali dialects. This would require the collection or curation of additional dialectal speech data and would further validate the scalability of our approach.
- **Self-Training with Pseudo-Labeling:** The robust dialectal ASR model developed in this work can be used to kickstart a virtuous cycle of improvement. It can be used to automatically transcribe large volumes of unlabeled dialectal audio data. These pseudo-labels, after filtering for confidence, can be added to the training set to re-train an even more powerful model. This self-training methodology could be a key strategy for overcoming the data scarcity bottleneck for non-standard dialects.[45]
- **Integration with Neural Language Models:** Future work should explore the fusion of our robust acoustic model with powerful external language models (e.g., n-gram models or Transformer-based LMs like BanglaBERT) during the decoding process. This can be achieved through techniques like shallow fusion or deep fusion, and is likely to yield significant improvements in WER by leveraging contextual linguistic constraints.

- **Cross-Dialectal Adaptation and Transfer Learning:** An interesting research question is to investigate the efficiency of adapting a model trained on one Bengali dialect to another, related dialect. This could potentially reduce the data and computation requirements for supporting new dialects, further enhancing the scalability of the system for a highly diverse linguistic landscape.

b3b1-ffbee4812aa6